\documentstyle[preprint,aps]{revtex}
\begin{document}
\def \sign{ \mathop{ \rm sign}\nolimits}
\def \trace{ \mathop{ \rm trace}\nolimits}
\draft
\title{Diagonal solutions to reflection equations in higher spin models }
\author{J. Abad and M. Rios}
\address{Departamento de F\'{\i}sica Te\'{o}rica, Facultad de Ciencias\\
Universidad de Zaragoza, 50009 Zaragoza, Spain}
\date{\today}
\maketitle
\begin{abstract}
A general fusion method to find solutions to the reflection equation in higher
spin representations starting from the fundamental one is shown. The method is
illustrated by applying it to obtaining the $K$ diagonal boundary matrices in
an alternating spin $1/2$ and spin $1$ chain. The hamiltonian is also given.
The applicability of the method to higher rank algebras is shown by obtaining
the $K$ diagonal matrices for a spin chain in the $\left\{ 3^* \right\}$
representation of $su(3)$ from the $\left\{  3\right\}$ representation.
\end{abstract}
Quantum integrable systems are usually treated by imposing periodic boundary
conditions. Recently there has been a growing interest in exploring other
possibilities compatible with integrability.
There is a method, proposed by Sklyanin  in the framework of the algebraic
Bethe ansatz
\cite{ri}
 and relying on previous results by Cherednik \cite{rmvi} , for obtain
integrable models with non-periodic boundary conditions.
Sklyanin's original formalism, which assumes the model is invariant under the
parity and time reversal symmetries,  was  extended to more general systems in
Refs. \onlinecite{rii,riii,riv} among others.
A careful analysis within the framework of  algebraic structures was carried
out in Refs. \onlinecite{rmi,rmii,rmiii}. For the $XXZ$ model, an alternative
approach, based on a diagonalization schema via vertex operators
\cite{rmvii}, can be found in Ref \onlinecite{rmviii}.

Sklyanin's formalism introduces a  boundary $K^{+} \left( \theta \right) $
matrix, which must verify the well-known reflection condition expressed by

\begin{eqnarray}
R(\theta-\theta') \left[  K^{+} (\theta) \otimes I\right] &&R(\theta+\theta')
\left[  K^{+} (\theta')  \otimes I\right]   \nonumber\\
&&=\left[  K^{+} (\theta ') \otimes I\right] R(\theta+\theta')
\left[  K^{+} (\theta ) \otimes I\right] R(\theta-\theta'). \label{ei}
\end{eqnarray}
Matrix $K^{+} \left( \theta \right) $ together with $R \left( \theta \right) $
determine the integrable system with open boundary conditions. In this way,
solutions to models associated to the fundamental representation of different
algebras have been found in Refs \onlinecite{rv,rvi,rmiv,rvii,rmv,rviii}

The concept of obtaining results in higher dimensional representations from
those in lower dimensional representations is known as fusion methods. For $R$
matrices it was worked out by Karowski \cite{rmzi}  and Kulish, Reshetikhin and
Sklyanin \cite{rmzii} and for $K$ matrices by Mezincescu and Nepomechie
\cite{rmziii} .

In this paper, we discuss the application of a fusion method to a
non-homogeneous  chain with spin $1/2$ and spin $1$ in alternating sites. The
hamiltonian is also given. As an example of extension of the method to higher
rank algebras, the method is applied to a chain based on the basic
representations of $su(3)$. The notation used is explained in figure 1.
\bigskip
\input epsf
\centerline{\epsfxsize=10cm  \epsfbox{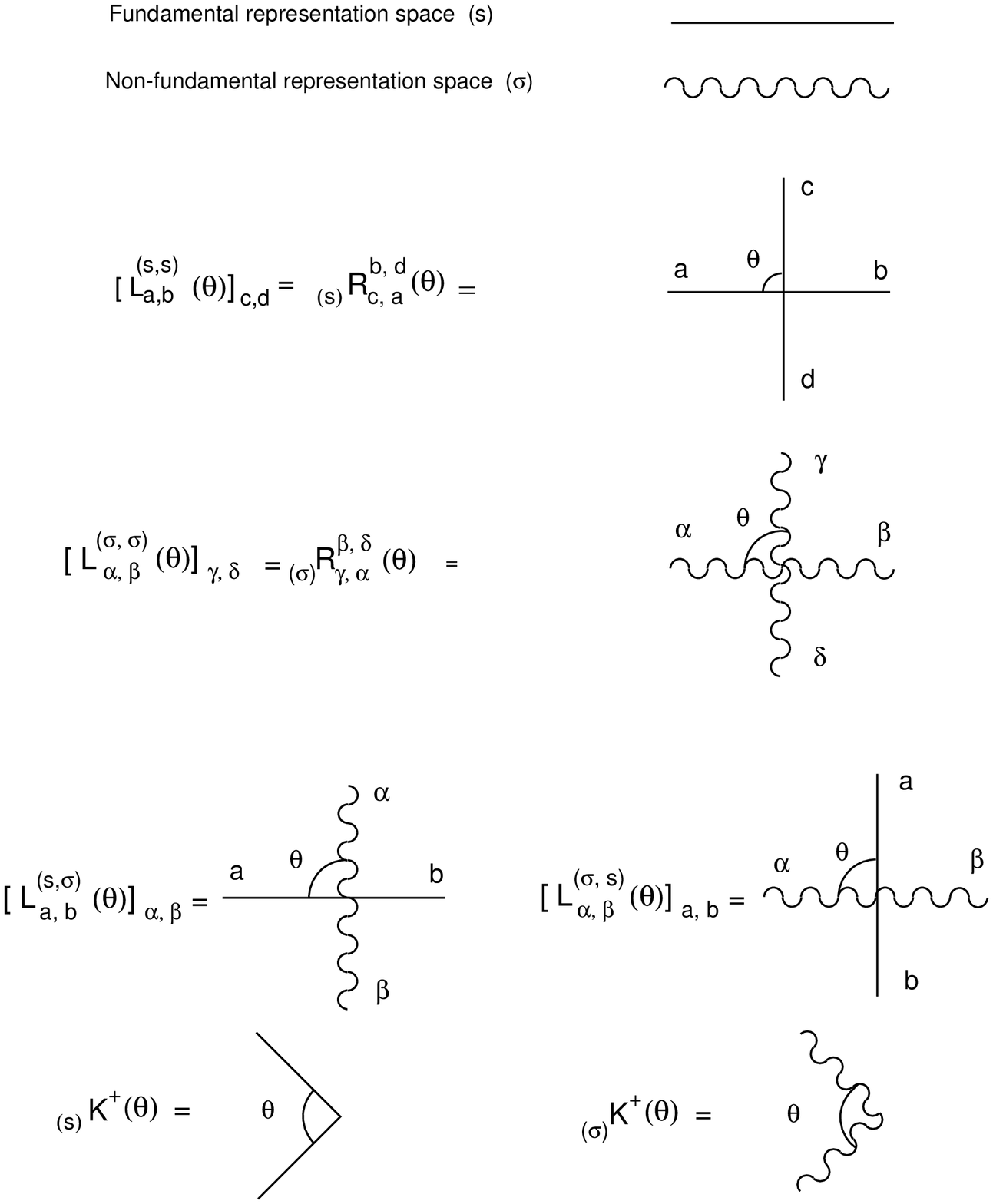}}
\centerline{Fig. 1}
\bigskip

In order to find the reflection matrices in a higher representation $\sigma$,
one should solve Eq. \ (\ref{ei}) with all operators in the required
representation. This would require the knowledge of the ${}_{\left( \sigma
\right)}R(\theta)$ operators, whose complexity increases with the dimension of
representation $d$, as they are represented by $d^2  \times d^2$ matrices.
Instead, we propose an alternative method.

Taking into account that Yang-Baxter equations have validity in any
representation, we can use them with either equal or different representations
for the auxiliary space and the site space. Thus, we distinguish two
representations  $s$ and $\sigma$ and consider the transfer matrices
\begin{equation}
T_{a,b}^{(s,s)} \left( \theta \right)  =
L_{a,a_{1}}^{(s,s)}\left( \theta \right)   \otimes L_{a_{1},a_{2} }^{(s,s)}
\left( \theta \right) \otimes \cdots \otimes
 L_{a_{N-1},b}^{(s,s)}\left( \theta \right)  ,
\label{eii}
\end{equation}
and
\begin{equation}
T_{\alpha,\beta}^{(\sigma,s)} \left( \theta \right)  =
L_{\alpha,\alpha_{1}}^{(\sigma,s)}\left( \theta \right)   \otimes
L_{\alpha_{1},\alpha_{2} }^{(\sigma,s)} \left( \theta \right) \otimes \cdots
\otimes
 L_{\alpha_{N-1},\beta}^{(\sigma,s)}\left( \theta \right)  .
\label{eiii}
\end{equation}
Following the method in ref. \cite{ri}, we build the doubled transition
operators
\begin{equation}
U_{a,b} ^{(s,s)} \left( \theta \right) = T_{a,c}^{(s,s)}  \left( \theta
\right){}_{(s)} K_{c,d}^{+} \left( \theta \right) T_{d,b}^{{(s,s)} ^{-1}
}\left(- \theta \right),
\label{eiv}
\end{equation}
and
\begin{equation}\label{ev}
U_{\alpha,\beta} ^{(\sigma,s)} \left( \theta \right) =
T_{\alpha,\gamma}^{(\sigma,s)}  \left( \theta \right){}_{(\sigma)}
K_{\gamma,\delta}^{+} \left( \theta \right) T_{\delta,\beta}^{{(\sigma,s)}
^{-1} }\left(- \theta \right).
\end{equation}
These expressions are represented in fig. 2.

\bigskip
\centerline{\epsfxsize=8cm  \epsfbox{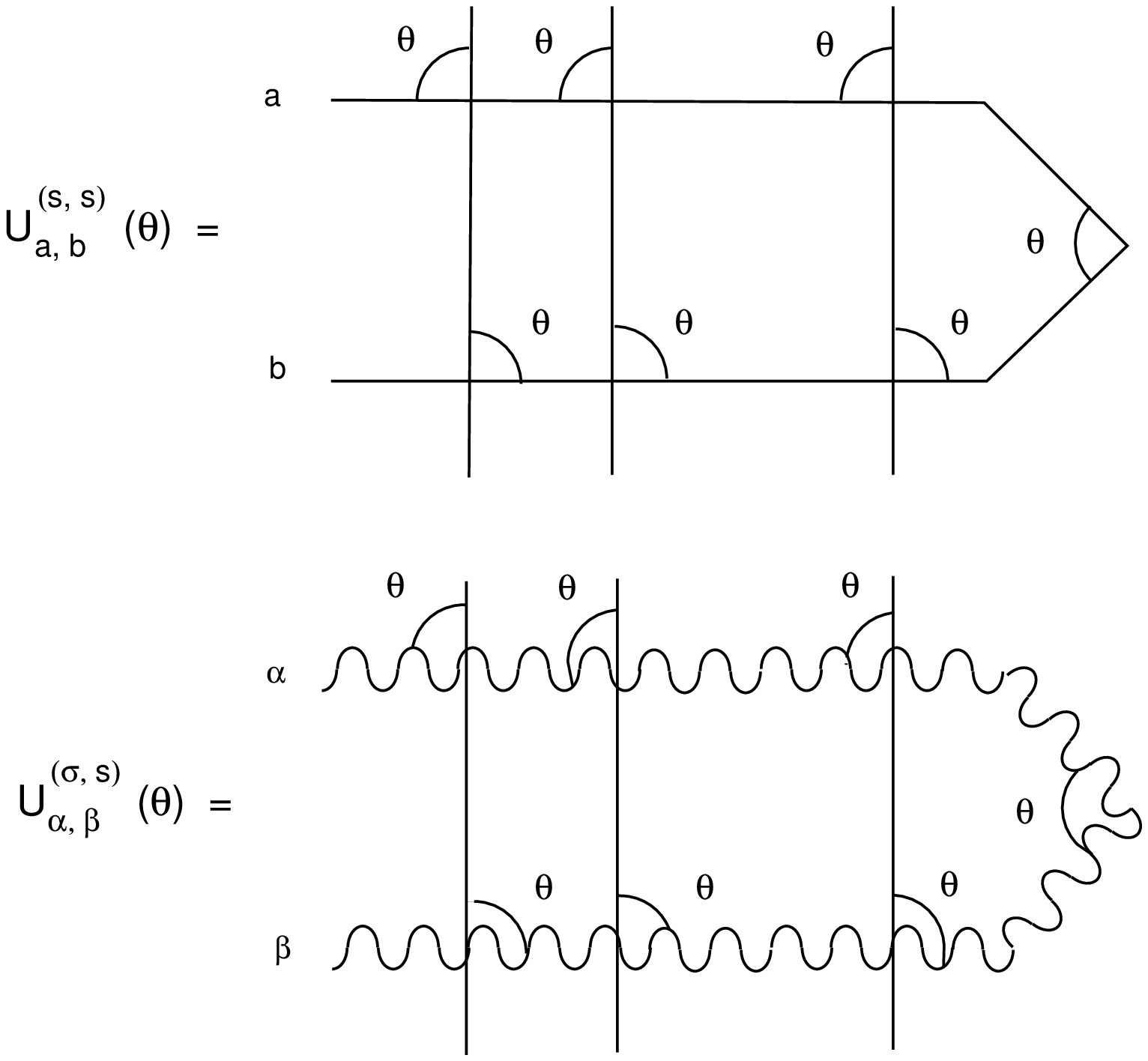}}
\centerline{Fig. 2}
\bigskip

The $U$ operators fulfill the Yang-Baxter equation

\begin{eqnarray}
L^{(\sigma,s)}\left( \theta -\theta'  \right)&&[ U^{ ( \sigma,s) } \left(
\theta \right)\otimes I  ]
L^{(s,\sigma)} \left( \theta +\theta'  \right)
[ U^{(s,s)}\left( \theta' \right) \otimes I  ]    \nonumber \\
 &&=[ U^{(s,s)}\left( \theta' \right)\otimes I ]
L^{(\sigma,s)}\left( \theta +\theta'  \right)
[ U^{(\sigma,s)}\left( \theta \right)\otimes I ]
L^{(s,\sigma)}\left( \theta -\theta'  \right) ,
\label{evi}
\end{eqnarray}

which is graphically expressed by fig. 3,

\bigskip
\centerline{\epsfxsize=10cm  \epsfbox{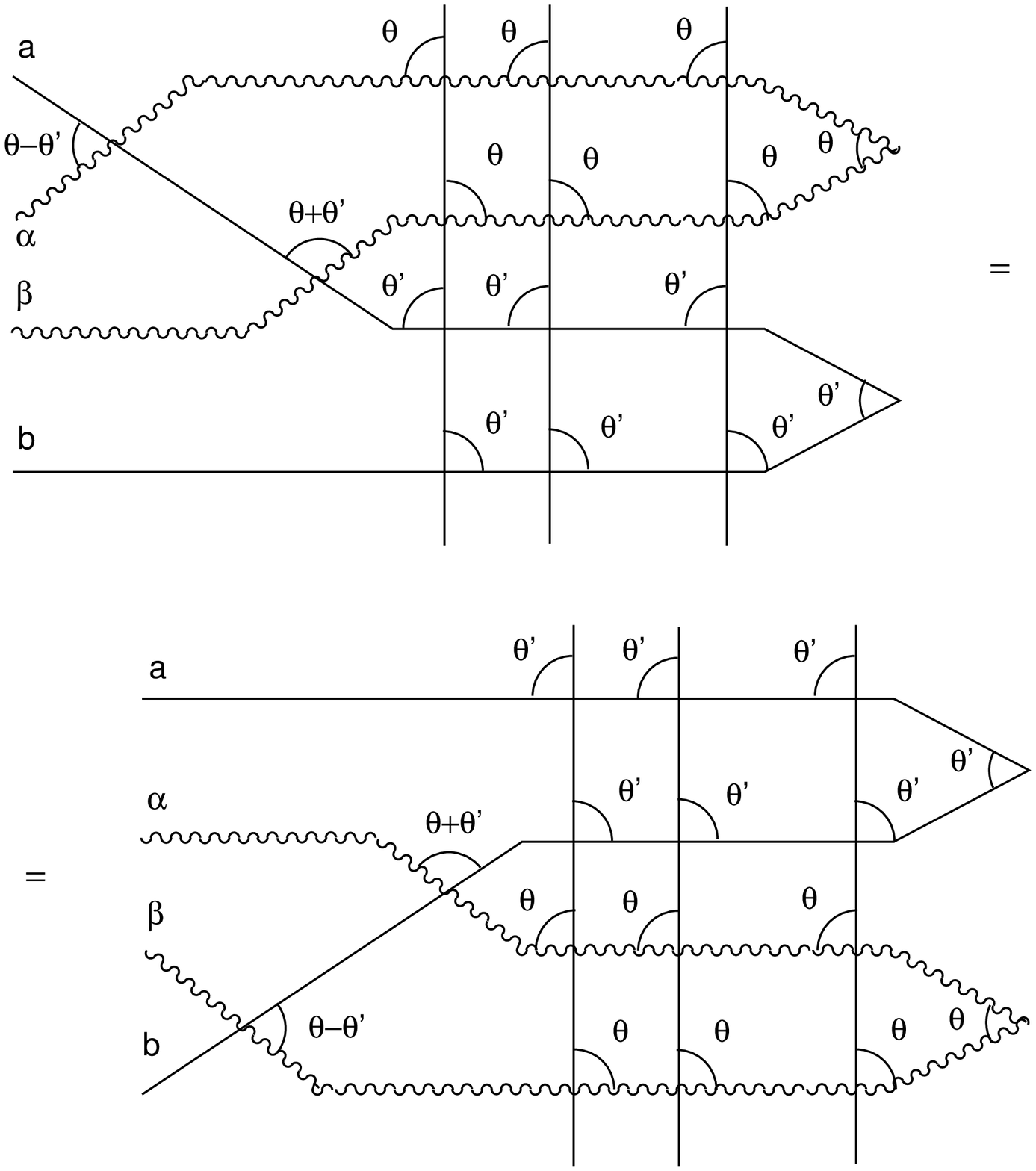}}
\centerline{Fig. 3}
\bigskip
\noindent whereas the reflection matrices $K$ must verify a relation similar to
(\ref{evi}), namely
\begin{eqnarray}\label{eviii}
L^{(\sigma,s)}\left( \theta -\theta'  \right)&&
[ { }_{(\sigma)} K^+\left( \theta \right)\otimes I  ]
L^{(s,\sigma)} \left( \theta +\theta'  \right)
[ { }_{(s)} K^+\left( \theta' \right)\otimes I  ]
  \nonumber \\
&&=[ { }_{\left(s \right)} K^+\left( \theta' \right)\otimes I  ]
L^{(\sigma,s)}\left( \theta +\theta'  \right)
[ { }_{(\sigma)} K^+\left( \theta \right)\otimes I  ]
L^{(s,\sigma)}\left( \theta -\theta'  \right) ,
\end{eqnarray}
which is graphically expressed by fig. 4

\bigskip
\centerline{\epsfxsize=10cm  \epsfbox{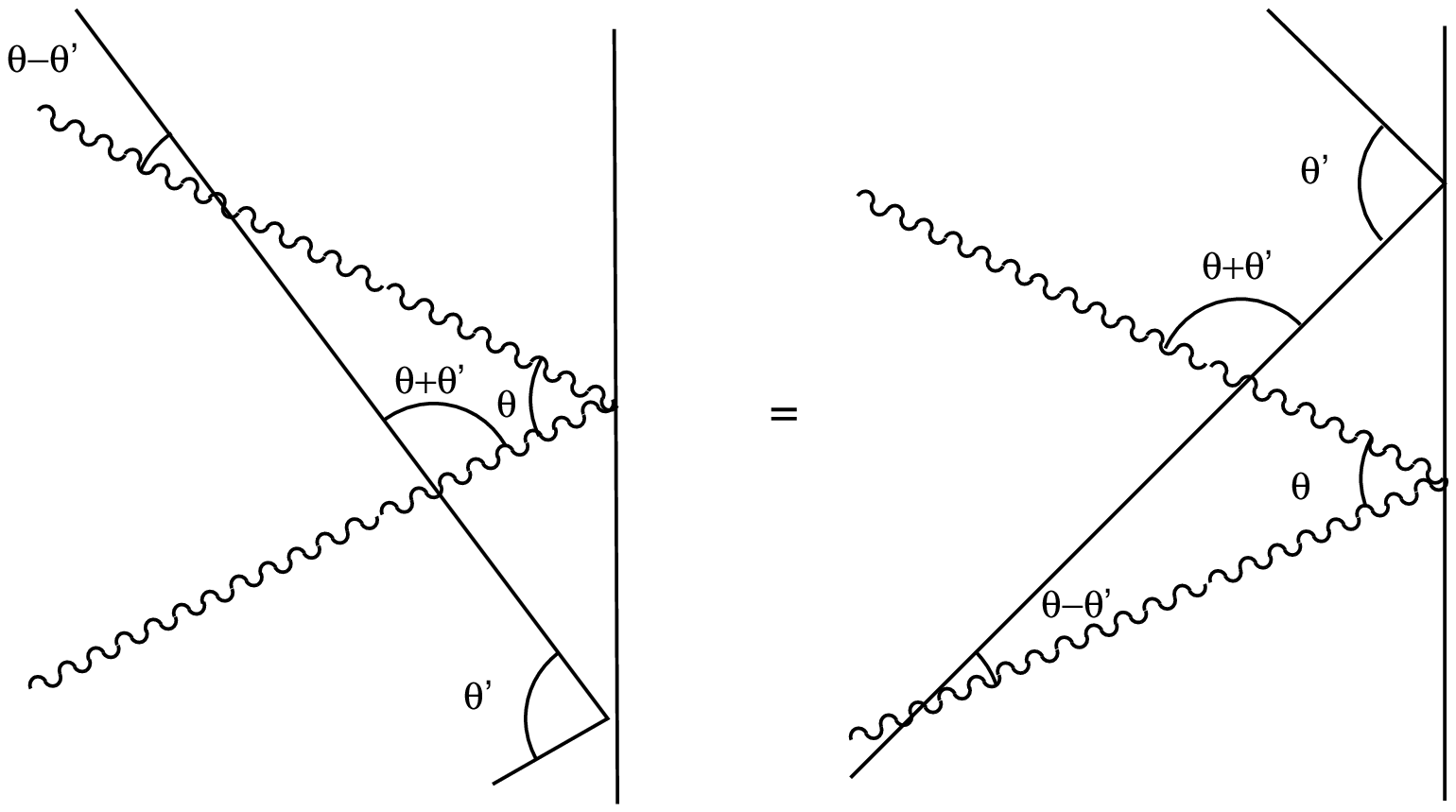}}
\centerline{Fig. 4}
\bigskip
The last equation provides the key for finding the matrices
${}_{(\sigma)}K^+(\theta)$ from the matrices in the fundamental representation
 ${}_{(s)}K^+(\theta)$ and the operators $L^{(\sigma,s)}$ and $L^{(s,\sigma)}$.

In order to illustrate the method, we apply it to $su(2)$. In this case we
start from the $XXZ$ model obtained in the fundamental representation of $U_q
\left( su(2) \right)$ and find the reflection matrices for spin $\sigma$.

The operator
\begin{equation}\label{eix}
L^{({1 \over 2},\sigma)}\left( \theta \right)=
\sinh{\left( \theta+{1 \over 2}\gamma\left( 1+2 \sigma^3\otimes S^3 \right)
\right)}+
\sinh{\gamma}\left( \sigma^+\otimes S^- +\sigma^-\otimes S^+ \right)
\end{equation}
where $q=\exp{\gamma}$, is already known \cite{rv}.
The operator $L^{(\sigma,{1 \over 2})}$  can be obtained from $L^{({1 \over
2},\sigma)}$ by transposition. The diagonal solution for $ {}_{({ 1\over
2})}K^{+}(\theta) $ can be found in \cite{rv} and is given by

\begin{equation}\label{ex}
{}_{({ 1\over 2})}K^{+} (\theta) =
\pmatrix{
K_+ \sinh{(\epsilon_+ -\theta)} & 0 \cr
0 & K_+ \sinh{(\epsilon_+ +\theta)}  \cr}
\end{equation}
$K^{+}$ and $\epsilon_{+}$ being arbitrary parameters.
By introducing (\ref{eix}) and (\ref{ex}) in (\ref{eviii}) we obtain $2 \sigma$
independent equations that determine the diagonal solution for the
${}_{(\sigma)}K^{+}(\theta)$ matrix.
\begin{equation}\label{exi}
{}_{(\sigma)}K_{i,j}^{+} (\theta)=
\delta_{i,j} K_{+}
\prod_{l=1}^{2\sigma}{\sinh{\left( \epsilon_{+}-(\sigma+{1 \over 2}-l )\gamma +
\sign(i-{1 \over 2}-l ) \theta \right)}
}
\end{equation}
As a particular case, we can take the spin $\sigma=1$, then the $K$-elements
are \cite{rmziv}
\begin{mathletters}
\begin{eqnarray}
{}_{(1)}K_{1,1}^{+}(\theta) =&&
K_{+} \sinh{(\epsilon_{+}-{\gamma \over 2}-\theta)}
\sinh{(\epsilon_{+}+{\gamma \over 2}-\theta)} \label{exii a} \\
{}_{(1)}K_{2,2}^{+}(\theta) =&&
K_{+} \sinh{(\epsilon_{+}+{\gamma \over 2}-\theta)}
\sinh{(\epsilon_{+}-{\gamma \over 2}+\theta)}  \label{exii b}\\
{}_{(1)}K_{3,3}^{+}(\theta) =&&
K_{+} \sinh{(\epsilon_{+}-{\gamma \over 2}+\theta)}
\sinh{(\epsilon_{+}+{\gamma \over 2}+\theta)}\label{exii c}
\end{eqnarray}
\end{mathletters}
The hamiltonian can be obtained from the $K$ and   ${}_{( \sigma)}R(\theta)$
matrices. For the $\sigma=1$ case, it is
\begin{eqnarray}
\label{exiii}
H=&&\sinh(\gamma)\sum_{i=1}^{N-1}\Bigl[
 \vec{S}_{i} \vec{S}_{i+1}-( \vec{S}_{i} \vec{S}_{i+1} )^2+
2 \sinh^2 (\gamma)\left( S_{i}^{z}  S_{i+1}^{z}-(S_{i}^{z}  S_{i+1}^{z})^2
\right)
 \nonumber\\
&&+2 \left( 1-\cosh (\gamma) \right) \left(  (S_{i}^{x}  S_{i+1}^{x} +S_{i}^{y}
 S_{i+1}^{y})
S_{i}^{z}  S_{i+1}^{z} + S_{i}^{z}  S_{i+1}^{z}  (S_{i}^{x}  S_{i+1}^{x}
+S_{i}^{y}  S_{i+1}^{y})  \right) \nonumber\\
&&+2 \sinh^2(\gamma)\left( (S_i^z)^2 +(S_{i+1}^z)^2 \right) \Bigl]\cr
&&+{ {\sinh^2(\gamma) \cosh(\gamma)}\over  \sinh{(\epsilon_{-}-{\gamma \over
2})}
 \sinh{(\epsilon_{-}+{\gamma \over 2})}  }
\left( \sinh(2 \epsilon_{-} ) S_1^z- \sinh(\gamma ) (S_1^z)^2
\right)\nonumber\\
&&+{ \sinh(\gamma)\over 2 \sinh{(\epsilon_{+}-{\gamma \over 2})}
 \sinh{(\epsilon_{+}+{\gamma \over 2})}  }
\left( \sinh(2 \epsilon_{+} ) S_N^z+ \sinh(\gamma ) (S_N^z)^2 \right) +ctn
\cdot I
\end{eqnarray}
which corresponds to the hamiltonian given in \cite{rix} and \cite{rx} with
boundary terms \cite{rmziv}.

For the isotropic spin ${1 /2}$  case ($XXX$ model), the diagonal solution of
the reflection equations is

\begin{equation} \label{exiv}
{}_{({1 \over 2})_{\rm {iso}}} K^{+}(\theta)=
\pmatrix{
\epsilon_{+}-\theta & 0 \cr
0 & \epsilon_{+}+\theta \cr
}.
\end{equation}
Using our procedure, we found the solution for spin $\sigma$
\begin{equation}\label{exv}
{}_{(\sigma)_{\rm{iso}}}K_{i,j}^{+} (\theta)=
\delta_{i,j} K_{+}
\prod_{l=1}^{2\sigma}{\bigl( \epsilon_{+}+\sigma+{1 \over 2}-k+
\sign(i-{1 \over 2}-k ) \theta\bigl)}.
\end{equation}
Of course, this result can also  be obtained from \ref{exi}  by replacing
$\gamma$ by  $-\eta$ and $\theta$ by $\eta \theta$ in the limit $\eta
\rightarrow 0$.

A very interesting application of the method is to a non-homogeneous chain
combining different kinds of spin in alternating sites. In reference
\cite{rmvi} a non-homogeneous chain based on the $su(2)$ algebra combining the
spin $s=1/2$ and $\sigma=1$ in alternating sites is described . We shall now
describe this chain with open boundary conditions. Following the same notation
as in \cite{rmvi} , we have

\begin{equation}\label{eni}
L^{({ 1\over 2},{ 1\over 2})}(\theta)=
\left (\matrix{
a & 0 & 0 & 0 \cr
0 & b & c & 0 \cr
0 & c & b & 0 \cr
0 & 0 & 0 & a \cr
}\right ), \qquad
L^{({ 1\over 2},1)}(\theta)=\left (\matrix{
a_1 & 0 & 0 & 0 & 0 & 0\cr
0 & b_+ & 0 &  c_1 & 0 & 0\cr
0 & 0 & b_- & 0 &  c_1 & 0\cr
0 & c_1 & 0 & b_- & 0 & 0\cr
0 & 0 &  c_1 & 0 & b_+ & 0\cr
0 & 0 & 0 & 0 & 0 & a_1\cr
}\right)
\end{equation}
with
\begin{mathletters}
\begin{eqnarray}
&&a(\theta)=\sinh{(\theta+\gamma)}, \label{enii a} \\
&&b(\theta)=\sinh{(\theta)}, \label{enii b} \\
&&c(\theta)=\sinh{(\gamma)}, \label{enii c} \\
\end{eqnarray}
\end{mathletters}
and
\begin{mathletters}
\begin{eqnarray}
&&a_1(\theta)=\sinh{(\theta+{3 \over 2}\gamma)}, \label{eniii a} \\
&&b_+(\theta)=\sinh{(\theta+{\gamma\over 2})}, \label{eniii b} \\
&&b_-(\theta)=\sinh{(\theta-{\gamma\over 2})}, \label{eniii c} \\
&&c_1(\theta)=\sinh{(\gamma)} \sqrt{2 \cosh{\gamma}}.  \label{eniii d}
\end{eqnarray}
\end{mathletters}
The monodromy matrix, using the representation $\sigma=1/2$ as auxiliary space,
is
\begin{equation}\label{eniv}
{}_{1 \over 2}T_{a,b}^{{\rm alt}}(\theta, \alpha)=
L^{{1 \over 2},{1 \over 2}}_{a,a_1}(\theta)\cdot
 L^{{1 \over 2},1}_{a_1,a_2}(\theta+\alpha) \cdots
L^{{1 \over 2},{1 \over 2}}_{a_{2N-2},a_{2N-1}}(\theta)
 L^{{1 \over 2},1}_{a_{2N-1},b}(\theta+\alpha).
\end{equation}

The doubled monodromy matrix can be calculated by using
reflection matrix (\ref{ex}) for spin 1/2
\begin{equation}\label{env}
{}_{1 \over 2}U_{a,c}^{{\rm alt}}(\theta)=
{}_{1 \over 2}T_{a,b}^{{\rm alt}}(\theta, \alpha) \cdot
{}_{1 \over 2}K_{c,d}^{+}(\theta)\cdot
{}_{1 \over 2}T_{d,b}^{{\rm alt}^{-1}}(-\theta).
\end{equation}
where, for purpose of simplicity, we have made $\alpha=0$.
Taking into account that
\begin{equation}\label{envic}
{}_{1 \over 2}K^{-}(\theta)={}_{1 \over 2}K^{{+}^t}(\theta-\gamma),
\end{equation}
we can obtain the hamiltonian through the $t$ operator. The explicit form for
the hamiltonian can be written by dividing it into different parts
\begin{equation}\label{envi}
H=\sum_{i=1 \atop i {\rm odd}}^{2N-3}{h_{(i, i+1, i+2)}^{(A)}}
\sum_{i=1 \atop i {\rm odd}}^{2N-1}{h_{(i, i+1)}^{(B)}}+ h_{(1)}^{C}+h_{(2N-1,
2N)}^{D}.
\end{equation}

The first two  parts are the usual hamiltonian \cite{rmvi}
 and the others are
introduced by the boundaries. After a long calculation, for the three-sites
operator we find
\begin{eqnarray}\label{envii}
h_{(i, i+1, i+2)}^{(A)}=&&
{1 \over 4}  \Bigl(\cosh{(\gamma)}-1\Bigr)\sigma+{1 \over 4}\cosh{(\gamma)}
 \Bigl(2\cosh{(\gamma)}-\cosh{(3 \gamma)} -1 \Bigr)\bar{\sigma} \nonumber\\
&&+\sinh ^2 {(\gamma)} \bigl[\cosh{(\gamma)}\bigr]^{{ 1\over 2}}cosh{({\gamma
\over 2})} U +4 \sinh ^2 {(2  \gamma)} \bar {U} \nonumber\\
&&+2\sinh{(\gamma)}\sinh{(2 \gamma)} W +2\sinh{({\gamma \over 2})}\sinh{(2
\gamma)}  \bigl[\cosh{(\gamma)}\bigr]^{{ 1\over 2}} V \nonumber\\
&&-{1 \over 2}\sinh^2{(\gamma)}\sigma \cdot S_z^2 +4 \sinh^2{(2\gamma)}
\bar{\sigma}
\cdot S_z^2 ,
\end{eqnarray}
where
\begin{eqnarray} \label{enviii}
\sigma=&&\sigma_x^{i}\cdot I \cdot \sigma_x^{(i+2)}+\sigma_y^{i}\cdot I \cdot
\sigma_y^{(i+2)}, \nonumber\\
\bar{\sigma}=&&\sigma_z^{i}\cdot I \cdot \sigma_z^{(i+2)}, \nonumber\\
U=&&I \cdot S_x^{i+1}\sigma_x^{(i+2)}+I \cdot
S_y^{i+1}\sigma_y^{(i+2)},\nonumber\\
\bar{U}=&&I \cdot S_z^{i+1}\sigma_z^{(i+2)},\nonumber\\
V=&&\sigma_x^{i}\cdot \{S_x , S_z\}^{(i+1)} \cdot \sigma_z^{(i+2)} +
\sigma_y^{i}\cdot \{S_y , S_z\}^{(i+1)} \cdot \sigma_z^{(i+2)} \nonumber\\
&&+\sigma_z^{i}\cdot \{S_x , S_z\}^{(i+1)} \cdot \sigma_x^{(i+2)} +
\sigma_z^{i}\cdot \{S_y , S_z\}^{(i+1)} \cdot \sigma_y^{(i+2)},\nonumber\\
W=&&\sigma_+^{i}\cdot S_-^{(i+1)^2} \cdot \sigma_+^{(i+2)} +
\sigma_-^{i}\cdot S_+^{(i+1)^2} \cdot \sigma_-^{(i+2)}.
\end{eqnarray}
The two-site operator is
\begin{eqnarray}\label{enix}
h_{(i,i+1)}^{(B)}=\sinh ^2{(\gamma )}  \Bigl[ \cosh{({\gamma \over
2})}&&\bigl[\cosh{(\gamma )}\bigr]^{{ 1\over 2}}  \bigl( \sigma_x^{i}\cdot
S_x^{(i+1)} +
\sigma_y^{i}\cdot S_y^{(i+1)} \bigr) \nonumber\\
&&+\cosh^2{(\gamma )} \sigma_z^{i}\cdot S_z^{(i+1)} +
\sinh ^2{(\gamma )} I^{(i)} \cdot S_z^{(i+1)^2}\Bigr]
\end{eqnarray}
and the two boundary terms
\begin{equation}\label{enx}
h_{(1)}^{(C)}={{\sinh^2{({3 \gamma \over 2})}} \over {4 \cosh{(\gamma)}}}
\sinh{(2 \gamma)}
\coth{(\epsilon_-)} \sigma_z^{(1)}
\end{equation}
and

\begin{eqnarray}\label{enxi}
h_{(2N-1, 2N)}^{(D)}=&& \sinh^2{(\gamma)}\coth{(\epsilon_+)}
\Bigl[   { {1-2 \cosh{(\gamma)} +\cosh{(3\gamma)}  } \over {4 \sinh{(\gamma)}
}}
 \sigma_z^{(2N+1)}       \nonumber\\
&& -\cosh{(\gamma)} \sinh{(\gamma)}  \bigl(  \sigma_z^{(2N-1)}\cdot
S_z^{(2N)^2}+
I \cdot   S_z^{(2N)} \bigr)    \nonumber\\
&&- \sinh{({\gamma \over 2})} \bigl[\cosh{(\gamma)} \bigr]^{{1 \over 2}}
\bigl(\sigma_x^{(2N-1)}\cdot \{S_x , S_z\}^{(2N)} +
\sigma_y^{(2N-1)}\cdot \{S_y , S_z\}^{(2N)} \bigr)
 \Bigr].
\end{eqnarray}
In this hamiltonian, the terms proportional to the identity have been removed.

In the alternating chain, we can build another system by taking the
representation $\sigma=1$ as auxiliary space. The doubled monodromy matrix
${}_{s}U$ is obtained with the reflection matrix ${}_1K$ given in (\ref{exii
a}-c) and the operators $L^{(1, 1)}$ and $ L^{(1, {1 \over 2})}$ which are
obtained  by transposing (\ref{eix}) . The new ${}_{1}t$ operator commutes with
the previous ${}_{1/2}t$ one, so it follows that both hamiltonians also
commute.

As an example of the application of the method to a higher rank algebra, we
will find the reflection matrices in a spin chain whose site space is in the
$\{3^*\}$ representation  of $su(3)$, starting from the chain whose site space
is in the other fundamental representation $\{3\}$.

We have two solutions in the fundamental representation $\{3\}$ given in
references  \cite{rvi} and \cite{rviii}. They are,
\begin{mathletters}
\label{exvi}
\begin{eqnarray}
{}_{\left\{ 3 \right\}}K_{1,1}^{+}(\theta) &&=
K^{+}  e^{\theta} \sinh{(\epsilon_{+}-{3 \over 2}\theta)} , \label{exvia}\\
{}_{\left\{ 3 \right\}}K_{2,2}^{+}(\theta) &&=
K^{+} \sinh{(\epsilon_{+}+{3 \over 2}\theta)} ,   \label{exvib}\\
{}_{\left\{ 3 \right\}}K_{3,3}^{+}(\theta) &&=
K^{+} e^{2 \theta}\sinh{(\epsilon_{+}+{3 \over 2}\theta)}   \label{exvic}
\end{eqnarray}
\end{mathletters}
and
\begin{mathletters}
\label{exvii}
\begin{eqnarray}
{}_{\left\{ 3 \right\}}K_{1,1}^{+}(\theta) &&=
K^{+} \sinh{(\epsilon_{+}-{3 \over 2}\theta)} , \label{exviia}\\
{}_{\left\{ 3 \right\}}K_{2,2}^{+}(\theta) &&=
K^{+}  e^{2\theta} \sinh{(\epsilon_{+}-{3 \over 2}\theta)}  , \label{exviib}\\
{}_{\left\{ 3 \right\}}K_{3,3}^{+}(\theta) &&=
K^{+} e^{ \theta}\sinh{(\epsilon_{+}+{3 \over 2}\theta)} . \label{exviic}
\end{eqnarray}
\end{mathletters}

By applying the method described above and knowing that
\begin{equation}\label{exviii}
L^{(\{3\},\{3^*\})} (\theta)=
\left (\matrix{
\bar{a} & 0 & 0 & 0 & \bar{c} & 0& 0 & 0 & \bar{d} \cr
0 & \bar{b} & 0 & 0 & 0 & 0 & 0 & 0 & 0 \cr
0 & 0 & \bar{b} & 0 & 0 & 0 & 0 & 0 & 0  \cr
0 & 0 & 0 & \bar{b}& 0 & 0 & 0 & 0 & 0  \cr
\bar{d} & 0 & 0 & 0 & \bar{a} & 0 & 0 & 0 & \bar{c}  \cr
0 & 0 & 0 & 0 & 0 &  \bar{b} & 0 & 0 & 0  \cr
0 & 0 & 0 & 0 & 0 & 0 &  \bar{b} & 0 & 0  \cr
0 & 0 & 0 & 0 & 0 & 0 & 0 &  \bar{b} & 0  \cr
\bar{c} & 0 & 0 & 0 & \bar{d} & 0 & 0 & 0 & \bar{a}  \cr
}\right ),
\end{equation}
with
\begin{mathletters}
\label{exix}
\begin{eqnarray}
&&\bar{a}(\theta)=\sinh{\left( {3 \over 2} \theta + {\gamma \over 2} \right)},
\label{exixa}\\
&&\bar{b}(\theta)=\sinh{\left( {3 \over 2} (\theta + \gamma ) \right)},
\label{exixb}\\
&&\bar{c}(\theta)=-\sinh{(\gamma)} e^{{{\theta +\gamma }\over 2}},
\label{exixc}\\
&&\bar{d}(\theta)=-\sinh{(\gamma)} e^{-{{\theta +\gamma }\over
2}},\label{exixd}
\end{eqnarray}
\end{mathletters}

we obtain, for the $\{3^*\}$ representation, the diagonal solutions
\begin{mathletters}
\label{exx}
\begin{eqnarray}
{}_{\left\{ 3^* \right\}}K_{1,1}^{+}(\theta) &&=
K^{+} e^{ \theta}\sinh{(\epsilon_{+}+{3 \over 2}\theta-{\gamma \over 2})},
\label{exxa}\\
{}_{\left\{ 3^* \right\}}K_{2,2}^{+}(\theta) &&=
K^{+}  e^{2\theta} \sinh{(\epsilon_{+}-{3 \over 2}\theta-{\gamma \over 2})},
  \label{exxb}\\
{}_{\left\{ 3^* \right\}}K_{3,3}^{+}(\theta) &&=
K^{+} \sinh{(\epsilon_{+}-{3 \over 2}\theta-{\gamma \over 2})}  \label{exxc}
\end{eqnarray}
\end{mathletters}

and
\begin{mathletters}
\label{exxi}
\begin{eqnarray}
{}_{\left\{ 3^* \right\}}K_{1,1}^{+}(\theta) &&=
K^{+} e^{ 2\theta}\sinh{(\epsilon_{+}+{3 \over 2}\theta+{\gamma \over 2})} ,
\label{exxia}\\
{}_{\left\{ 3^* \right\}}K_{2,2}^{+}(\theta) &&=
K^{+}  \sinh{(\epsilon_{+}+{3 \over 2}\theta+{\gamma \over 2})},
  \label{exxib}\\
{}_{\left\{ 3^* \right\}}K_{3,3}^{+}(\theta) &&=
K^{+} e^{ \theta}\sinh{(\epsilon_{+}-{3 \over 2}\theta+{\gamma \over 2})} .
\label{exxic}
\end{eqnarray}
\end{mathletters}

The method  is quite general and will be applied in future work to look for
non-diagonal solutions with arbitrary spin and to models based on higher rank
algebras.

\acknowledgements

We are grateful to Professor R. Nepomechie for his very useful comments
and to Professor J. Sesma for the careful reading of the manuscript. This work
was partially supported by the Direcci\'{o}n General de Investigaci\'{o}n
Cient\'{\i}fica y T\'{e}cnica, Grant No PB93-0302 and AEN94-0218

\end{document}